\documentclass[12pt]{article}
\usepackage{amsfonts,epsfig}
\usepackage{amsmath,amsfonts}
\usepackage{amssymb,graphicx}
\def\na{\nabla}
\def\pa{\partial}
\def\L{\Lambda}

\def\k{\kappa}

\def\d{\delta}
\def\h{\hat}
\def\b{\bar}
\def\t{\tilde}
\def\f{\frac}

\def\l{\label}

\def\a{\alpha}

\def\la{\lambda}
\def\g{\gamma}
\def\G{\Gamma}
\def\m{\mu}
\def\n{\nu}
\def\na{\nabla}
\def\r{\rho}
\def\s{\sigma}
\def\c{\check}
\def\be{\begin{equation}}
\def\ee{\end{equation}}
\def\ba{\begin{eqnarray}}
\def\ea{\end{eqnarray}}

\begin{document}
\vspace{10pt}
\begin{center}
  {\LARGE \bf Graviton emission from the brane in the bulk 
\\\vspace*{0.5cm} inextra dimension } \\
\vspace{20pt}
Mikhail Z.Iofa\\
\vspace{15pt}

\textit{Skobeltsyn Institute of Nuclear Physics,
Moscow State University, Moscow,119992 , Russia}\\
\end{center}

\vspace{1 cm}
\begin{abstract}
In a model of 3-brane embedded in 5D space-time we calculate the graviton 
emission from the brane to the bulk.
Matter is confined to the brane, gravitons produced in reactions of matter on the brane 
escape to 
the bulk. The Einstein equations which are modified by the terms due to graviton production are 
solved perturbatively, the leading order being that without the graviton production. In the period of 
late cosmology, in which in the generalized Friedmann equation the term linear in the energy density 
of matter in dominant, we calculate the spectrum of gravitons (of the tower of 
Kaluza-Klein states) and the collision integral in the Boltzmann equation. We find the energy-momentum 
tensor of the emitted gravitons in the bulk, and using it show that corrections due to graviton 
production  to the 
leading-order terms in the Einstein equations are  small, 
and the perturbative approach is justified.   
We calculate the difference of abundances of ${}^4 He$ produced in primordial 
nucleosynthesis in the models with and without the graviton production, and 
find that the difference is a very small number, much smaller than that estimated previously.
\end{abstract}



\section{Introduction}

Brane-world scenarios with a 3-brane  identified with the observable 
Universe which is embedded in a higher-dimensional space-time provide an alternative 
to the standard 4D cosmology \cite{BDL1,BDL2,kraus,col,maeda}, reviews \cite{rubrev,maart,durrer}.
A necessary requirement on these 
models is that they should reproduce  
 the main observational cosmological data, the age of the 
Universe, abundances of elements produced in primordial nucleosynthesis, etc.
A general property of the models with extra dimensions is that gravity 
propagates in the extra dimensions independent of whether the ordinary matter is 
confined to the brane or not. This entails a peculiar property of the models 
with extra dimensions which is absent in the standard cosmology: gravitons 
which are produced in reactions of particles of  
 matter on the brane can escape from the brane and propagate in the bulk 
\cite{tan,heb,lan1,lan2,rusdur,kir}. 
As a 
consequence the Einstein equations contain terms accounting for the 
graviton emission. Cosmological evolution of matter on the brane is also 
affected by this process. 

Roughly the energy loss on the brane due to the process $a+b\rightarrow G +X$ can 
be estimated as 
$$
\f{d\h{\r}}{dt}=-<n_a n_b \s_{a+b\rightarrow G +X} v E_G>
,$$  
where in the radiation-dominated period of the evolution of the Universe $n_a ,n_b \sim T^3$ and 
$ E_G \sim T$. This yields 
$$
\f{d\h{\r}}{dt}\sim -\k^2 T^8
.$$
Here $T$ is temperature of the Universe and $\k^2 =8\pi/M^3$, where $M$ is the 5D Planck mass,
is the 5D gravitational constant. 
Although at the late stages of the cosmological evolution the energy loss to the bulk is 
small, at high temperatures this effect can be significant.

In the present paper we consider the problem of graviton production and emission 
to the bulk in a model with one 3-brane embedded in the bulk with one  
extra dimension. We perform our calculations in a picture in which the metric is 
time-dependent and the brane is 
located at a fixed position in the extra dimension.

The Einstein equations including the terms due to graviton production are solved in the perturbative 
approach. In the leading order we neglect 
the graviton production, and include it in the next order. 
The classical background metric is the leading-order solution of the 5D Einstein equations. 
Background metric is a 
warped extension of the metric on the 
brane to the bulk with the warping factor $e^{\pm\m
y}$ \cite{BDL2}. Here $\m\sim\sqrt{-\Lambda}$, where $\Lambda/\k^2$ is the 5D cosmological constant 
and $y$ a coordinate of the extra dimension. 

We perform  calculations in the period of late cosmology. In this period in 
the generalized Friedmann equation
$$
H^2 (t) =\f{\m\k^2\h{\r}}{3}+\left(\f{\k^2\h{\r}}{6}\right)^2 +\ldots
$$ 
the term linear in
 energy density on the brane is much larger than the 
quadratic 
term, or, equivalently, $\m\gg\k^2\h{\r}$.
For the value of $\m$ of order $10^{-12} GeV$, which we adopt for
numerical estimates, the range of temperatures of the Universe characteristic to
the period of late cosmology extends to $T\sim 5\cdot 10^2 GeV$.

Evolution of the energy density on the brane $\h{\r}$ is determined 
 by the Boltzmann equation. The collision term in the Boltzmann 
equation accounts  for the graviton emission from the brane. To calculate the collision term, solving
 the equations of motion for fluctuations over the background metric, 
 we find the spectrum and the eigenfunctions of of the tower of  Kaluza-Klein 
gravitons and the energy-momentum tensor of the emitted gravitons.

Solution of the Einstein equations
 requires integration of the 5D energy-momentum tensor over 
coordinates of the extra dimension.  This, in turn, requires determination of the form of 
the 5D energy-momentum tensor not only on the brane, where gravitons are emitted, but also in the 
bulk.  
Integrating the geodesic equations for the null geodesics 
along which gravitons propagate  and using 
conservation equations for the energy-momentum tensor in the 
bulk, we find the form of the energy-momentum tensor in the bulk. 

Using the obtained energy-momentum tensor in the bulk, we show that in the period of late cosmology 
in the Friedmann equation corrections  to the leading-order terms 
 due to graviton emission are small, 
and the  perturbative approach is consistent.

Graviton emission changes cosmological evolution of matter on the brane. Time  (temperature)
 dependence of of the Hubble function determined from the Friedmann equation which includes the terms due 
to graviton emission is different from that  in the standard cosmological model. This, 
in turn, results in
a change of abundances of light elements produced in primordial nucleosynthesis \cite{steig}.
We find that the difference of  abundances of ${}^4 He$ produced in primordial 
nucleosynthesis calculated in the models with and 
without the 
graviton production  is a  small number, much smaller than that estimated in
\cite{heb,lan1,lan2}.  A crude estimate of the graviton
production in the early
cosmological period, which takes ito account the bounce of produced gravitons back to the brane
\cite{lan2}, 
does not alter this result.


\section{System of Einstein equations}
We consider the 5D models with one 3D brane embedded in the
bulk with the action
\be
\l{1}
S_5  =\f{1}{2\k^2}\left[\int\limits_\Sigma \,d^5 x\sqrt{-g^{(5)}}
(R^{(5)} -
2\L) + 2\int\limits_{\pa {\Sigma}}K \right] -\int\limits_{\pa 
{\Sigma}}
\,d^4 x\sqrt{-g^{(4)}} \,\h{\s} -
\int\limits_{\pa
{\Sigma}}d^4 x \sqrt{-g^{(4)}}L_m ,
\ee
where $x_4 \equiv y$ is coordinate of the infinite extra dimension,
 $\k^2 =8\pi/M^3$.   
 We consider a class of metrics of the form
\footnote{The indices $i,j$ run over 0,...,4, the Greek indices are
0,...,3, and
$a,b =1,2,3$ }
\be
\l{m}
ds^2_5 =g^{(5)}_{ij}dx^i dx^j =
-n^2 (y,t)dt^2 + a^2 (y,t)\eta_{ab} dx^a dx^b +dy^2 \equiv dy^2 +g_{\m\n}dx^\m 
dx^\n.
\ee
The brane is spatially flat and is located at the fixed position $y=0$.
The 5D Einstein equations are
\be
\l{1a}
G_{ij}\equiv R_{ij} -\f{1}{2}g_{ij}R =\k^2 T^{(5)}_{ij}-g_{ij}\Lambda 
-\d_{ij}^{\m\n}\f{\sqrt{-g^{(4)}}}{\sqrt{-g^{(5)}}}\d (y)g_{\m\n}\k^2\h{\s}
.\ee
Here $T^{(5)}_{ij}$ is the sum
of the energy-momentum tensor of
 matter confined to the brane  and the bulk energy-momentum tensor.
The energy-momentum tensor of matter on the brane  is taken
in the form
\be
\l{6a}
\d (y)\tau_\m^\n=diag\,\d (y)\{-\h{\r},\h{p},\h{p},\h{p}\}.
\ee
The  components of the Einstein
tensor $G_{ij}$ 
\ba
\l{5a0}
&{}&G_{00}=3\left[\f{\dot{a}^2}{a^2}-
n^2\left(\f{a''}{a}+\f{{a'}^2}{a^2}\right)\right]
\\\l{5a5}
&{}&G_{44}=3\left[\left(\f{{a'}^2}{a^2}+\f{a'
n'}{a\,n}\right)-\f{1}{n^2}\left(\f{\dot{a}^2}{a^2}-
\f{\dot{a}\dot{n}}{a\,n}+\f{\ddot{a}}{a}\right)\right]
\\\l{5}
&{}&G_{04} =3\left(\f{n'}{n}\f{\dot{a}}{a}-\f{\dot{a'}}{a}\right)
\ea
satisfy the relations
\footnote{Prime and dot denote differentiation over $y$ and $t$.}
(cf. \cite{BDL2})
\ba
\l{2a}
G_0^0-G_4^0 \f{\dot{a}}{a'}=\f{3}{2a' a^3}F'
\\
\l{3a}
G_{44}-G_{04}\f{a'}{\dot{a}}=\f{3}{2\dot{a} a^3}\dot{F},
\ea
where
\be
\l{4a}
F=(a' a)^2 -\f{(\dot{a} a)^2}{n^2}
\ee
In the leading order
 we neglect emission of the gravitons produced in
collisions of particles of matter on the brane in the bulk. In the next 
order we take into account the graviton emission into the bulk, 
 the energy-momentum tensor in the bulk is  $\c{T}^i_j$.  

For the following it is convenient to introduce
the normalized expressions for energy
density, pressure and
cosmological constant on the brane which all have the same
dimensionality $[GeV]$
\be
\l{8a}
\m =\sqrt{-\f{\L}{6}},\quad \s =\frac{\k^2\h{\s}}{6},
\quad \r =\frac{\k^2\h\r}{6}, \quad {p} =\frac{\k^2\h{p}}{6}.
\ee
The functions $a(y,t)$ and $n(y,t)$ satisfy the  junction conditions on the
brane \cite{BDL2} 
\footnote{We assume invariance $y\leftrightarrow -y$.}
\ba
\l{9a}
&{}& \f{a' (0,t)}{a(0,t)}= -\s -\r(t)
\\\nonumber
&{}& \f{n' (0,t)}{n(0,t)} =2\r (t) +3p (t) -\s
\ea
Using reparametrization of $t$, we set $n(0,t)=1$, i.e. $t$ is the proper time on the brane.
Eq. (\ref{3a}) can be rewritten as
\be
\l{10a}
\dot{F}= -\m^2 \dot{(a^4 )} -\f{\k^2}{6}(a^4 )' \c{T}_{04} 
+\f{\k^2}{6}\dot{(a^4)} \c{T}_{44}
.\ee
On the brane, using junction conditions and setting $\s\simeq\m$, we have 
\footnote{From the 5D Einstein equations and the fit of the data on abundance of 
${}^4 He$ produced in nucleosynthesis it follows that $\s^2 =\m^2(1 +O(H^2_0  
/\m^2 ))$, where $H_0$ is the present-time Hubble parameter 
\cite{mi1,mi2}. For $\m \sim 10^{-12} GeV$ correction is $\sim10^{-60}$. } 
\be
\l{11a}
\dot{F}= \m^2 \dot{(a^4 )} +\f{2\k^2 a^4}{3}(\r +\m ) \c{T}_{04}
+\f{2\k^2 a^3\dot{a} }{3} \c{T}_{44}
.\ee
 Integrating (\ref{11a}) in the
interval $(t,\, t_l )$, where the initial time $t_l$  (of the onset of 
the period of the 
late cosmology) is defined below,  we obtain
\be
\l{12a}
F(0,t)=\m^2 a^4 (0,t) +\f{2\k^2}{3} \int\limits_{t_l}^t dt'
\c{T}_{04}(t' )(\r (t') +\m){a^4 (0,t')}
+ \f{2\k^2}{3}\int\limits_{t_l}^t dt'\c{T}_{44}(t' )\dot{a}(0, t' ) a^3 (0,t' ) 
-C.
\ee

Substituting expression (\ref{4a}) for $F$ and using the junction conditions,
we rewrite (\ref{12a}) in a form of the generalized Friedmann equation
(cf. \cite{BDL1,BDL2})
\be
\l{14a}
H^2 (t) = \r^2 (t) +2\m\r (t) +\m \r_w (t)
-\f{2\k^2}{3 a^4 (0,t)}\int\limits_{t_l}^t dt'\left[
\c{T}_{04}(t' )(\r (t') +\m)+\c{T}_{44}(t' )H(t' )\right] a^4 (0,t')
.\ee
Here we introduced
\be
\l{13a}
 H(t) =\frac{\dot{a}(0,t)}{a (0,t)},\qquad \r_w (t) =\r_{w0}\left(\f{a(0,t_0 )}{a(0,t)}\right)^4
\ee
where $\r_w (t)$ is the so-called Weyl radiation term \cite{maart}, $\m\r_{w0}=C/a^4 (0,t_0 )$.
To comply with the observational data, the Weyl radiation term, which appears in the leading-order 
solution to the Einstein equations should be small
\footnote{The Weyl radiation term $\r_w$ is small as compared to the
radiation energy density
$\r_r$. BBN
constraints to the ratio of the Weyl radiation term to the photon energy density at the temperatures
$T\lesssim 0.8 MeV$ are
$-0.4<\r_w/\r_\g <0.1$ \cite{ichiki}, or $|\r_w/\r_\g |<0.55$ \cite{mi2}.}. 
Assuming that the terms with the 
bulk 
energy-momentum tensor can be treated perturbatively, we neglect corrections to the Weyl radiation term  
due to graviton emission from the brane. 

Substituting from the junction conditions
the expressions for $n'/n$ and $a'/a$ at $y=0$,
we transform the (04) component of the
Einstein equations (\ref{5}) to the form
\be
\l{22a}
\dot{\r} +3H(\r +p )=\f{\k^2 \c{T}_{04}}{3}
.\ee
On the other hand, the same equation, which is a generalization of the 
conservation equation for the
energy-momentum tensor of the matter confined to the brane to the case with the
energy-momentum flow in the bulk, is 
obtained by integration across the brane 
of the 5D  conservation law
$\nabla_i T^i_0 =0$ \cite{tan}. 

Some relevant equations following from general approach to the 3-brane embedded in the AdS bulk
are found in the appendix A.

\section{Period of late cosmology}

 Graviton production by hot matter is sufficiently intensive in the radiation-dominated period 
of cosmology.
In the period of late cosmology the terms linear in energy density are 
dominant in the Friedmann equation. Equivalently, in the radiation-dominated 
period of late cosmology, this can be stated as  
$\r_r(t)/\m <1$, or
\be
\l{2ca}
\f{\r_r (T)}{\m} =\f{\k^2 \h{\r}_r (T)}{6\m}=
\f{4\pi^3 g_* (T) T^4}{90 \m M^3} \simeq \f{4\pi^3 g_* (T) T^4}{90(\m M_{pl} )^2}<1
,\ee
where we used that $\m M^2_{pl}/M^3\simeq 1$ \cite{mi1,mi2}. Taking $\m \sim 10^{-12} 
GeV$, we find that the approximation of late cosmology is valid up to the temperatures 
of order $5\cdot 10^2 GeV$.

In the leading  approximation, in which the bulk energy-momentum tensor is set to zero, 
 from the system of the 5D Einstein equations 
and junction conditions on the brane is obtained the unique extension of the metric the components 
from the brane to the bulk \cite{BDL2}
\ba
\l{21b}
&{}&a^2 (y,t) =\f{a^2 (0,t)}{4}\left[e^{2\m |y|} \left(\left( \f{\r}{\m}\right)^2 +
\f{\r_w}{\m}\right)+ e^{-2\m |y|} \left(\left( \f{\r}{\m} +2\right)^2  +\f{\r_w}{\m}\right)
\right.\\\nonumber-
&{}&\left.2\left(\f{\r}{\m}\left( \f{\r}{\m} +2\right) +\f{\r_w}{\m}\right)\right]
,\ea
and
\be
\l{21c}
n(y,t)= \f{\dot{a}(y,t)}{\dot{a}(0,t)}
.\ee

Without the Weyl radiation term the function  $a^2 (y,t)$ is \footnote{The expression for 
$a^2 (y,t)$ in \cite{BDL2} contained no brane 
tension $\s$ .}
\be
\l{21a}
a^2 (y,t) =\f{a^2 (0,t)}{4}\left[e^{2\m |y|}
\left(\f{\r}{\m}\right)^2 +
e^{-2\m |y|} \left(\f{\r}{\m}+2\right)^2 -
2\f{\r}{\m}\left(\f{\r}{\m}+2\right) 
\right]
,\ee
where $a(0,t)$ is a solution of the Friedmann equation on the brane.
$a^2 (y, t)$  (\ref{21a}) As a function of $y \,\,a^2 (y, t)$  (\ref{21a}) has the minimum 
equal to zero at the point $|\b{y}|$
\be
\l{mp}
e^{2\m |\b{y}|} =1+\f{2\m}{\r}.
\ee
The function $n(y,t)$ has no zeroes.
At the point $\b{y}$ the scalar curvature $R^{(5)}$ is finite. 
The function (\ref{21b}) has no zeroes, but  the 
function $n(y,t)$ has a zero. Again this zero is a coordinate singularity.

In the region $0<|y|<|\b{y}|$ and for $\r/\m\ll 1$ the functions $a(y,t)$ and $n(y,t)$ can be 
approximated as
\ba
\l{22}
&{}& a(y,t)\simeq a(0,t)e^{-\m|y|}\\\nonumber
&{}& n(y,t)\simeq e^{-\m|y|}.
\ea
With (\ref{22}) we obtain the approximate form of the metric in 
the region $0<|y|<|\b{y}|$
\be
\l{24}
ds^2\simeq dy^2 +e^{-2\m|y|}(-dt^2 + a^2 (0,t) \eta_{ab}dx^a dx^b )
.\ee
With the approximate metric we have $a'(0,t)/a(0,t)=-\m $, which for $\r/\m\ll 1$ is 
close to the exact junction condition $a'(0,t)/a(0,t)=-(\m +\r )$.

In the radiation-dominated period
solution of the Friedmann equation
\be
\l{1c}
H^2 =2\m\r +\r^2    
\ee
for the radiation energy density is
\be
\l{25}
\r (t) =\f{1}{8\m t^2 +4 t}.
\ee
In the period of late cosmology the leading term is
$$
\r (t) \simeq\f{1}{8\m t^2}.
$$ 
If $\r(t)\gg \m$, we have $\r (t)\simeq 1/4t$.

\section{Fluctuations of background metric}


To calculate the spectrum of the Kaluza-Klein tower of gravitons, we solve 
the field equations for fluctuations  $h_{ij}$ over the
background metric $g_{ij}$. The purpose of this section is to clarify the issue of
the gauge conditions in solving the equations for fluctuations. We show that the traceless transverse conditions
on $h_{ij}$ cannot be imposed as gauge conditions at the level of the action, but can be imposed on solutions of 
the equations of motion.

 The part of the action 
quadratic 
in fluctuations is
\ba
\l{2f}
I =\f{1}{2}\int\sqrt{-g^{(5)}}\left[
(R -\L_{(5)} ) \left(-\f{1}{2}h_i^j h_j^i +\f{1}{4}h^2\right)-
R_i^j h_j^i h +2 R_i^j h_j^k h_k^i \right.\\\nonumber
\left.
+ \f{1}{2}\left( 2h_{qi;k}h^{ik;q} -h_{ik;q}h^{ik;q} +h_{,q}h^{,q}
-2h_{,i} h^{ik}{}_{;k} \right)\right]
.\ea
The action $I$ is invariant under the gauge transformations
\be
\l{gt}
\t{h}_{kl} =h_{kl} -(\na_k \xi_l +\na_l \xi_k ),
\ee
where $\na$ is defined with respect to the metric $g_{ij}$.
With the background metric (\ref{m}), $ds^2= dy^2 +g_{\m\n}dx^\m
dx^\n$, the gauge transformations
(\ref{gt}) take the form
\ba
\l{gt4}
 &{}&\t{h}_{44} =h_{44}-2\pa_4{\xi_4}\\
\l{gtm}
&{}&\t{h}_{4\m} =h_{4\m}-(\pa_\m\xi_4 +\pa_4 \xi_\m
-2\Gamma^\n_{\m 4}\xi_\n )\\
\l{gtmn}
&{}&\t{h}_{\m\n} =h_{\m\n}-(D_\m\xi_\n +D_\n \xi_\m
-2\Gamma^4_{\m\n}\xi_4 )
,\ea
where $D_\m$ is defined with respect to the metric $g_{\m\n}$.

To solve the equations following from the action (\ref{2f})  in the 
region $0<y<\b{y}$, 
instead of the exact metric, we  use the
approximate metric (\ref{24}). In the region $0<y<\b{y}$   we rewrite it 
as
\be
\l{ap}
ds^2 = dy^2 +b(y)\b{g}_{\m\n}(x)dx^\m dx^\n
,\ee
where $b(y)=e^{-2\m |y|}$.
We impose the gauge conditions $h_{4i}=0$.
The gauge condition $\t{h}_{44}=0$ can be realized by performing transformation with
\be
\l{gc1}
\xi_4 (y, x^\m ) =\f{1}{2}\int^y_{-y} dy' h_{44} (y' , x^\m )
\ee
Condition $\t{h}_{4\m} =0$ is imposed by taking
\be
\l{gc2}
\xi_\m (y, x^\r ) = b(y)\left(C_\m (x^\r ) +\int^y_{-y} dy' b^{-1} (y')
(h_{4\m}-\pa_\m \xi_4  )(y' ,x^\r ) \right)
.\ee
 $C_\m (x)$  are arbitrary functions.

From the Einstein equations for the background metric (\ref{1a}) follow the 
relations
\ba
\l{3f}
&{}& R= \f{10}{3}\Lambda -\f{2}{3}\k^2 \d (y)(\tau_\m^\m-4\h{\s}) \simeq 20\m^2 
+16\m 
\d(y)\\
&{}& R_\m^\n =
\f{1}{2}\d^\n_\m (R-2\L )+\k^2 \d (y)(\tau_\m^\n -\h{\s}\d_\m^\n )
 \simeq\d^\n_\m (-4\m^2 +2\m  \d (y) ) +\k^2\d(y) \tau_\m^\n,
\ea
where in the radiation-dominated period we have set $\tau^\m_\m =0$.  
The Ricci tensor calculated with the metric (\ref{ap}) is
\be
\l{4f}
R_\m^\n  = -\f{\d_\m^\n}{2}\left(\f{b''}{b} +
\f{{b'}^2}{b^2}\right)+\b{R}_\m^\n (\b{g}) = \d_\m^\n (-4\m^2 +
2\m  \d (y)  )+ {R}_\m^\n (\b{g})
,\ee
where we used the relations $b'/b=-2\m\, \mbox{sign} (y)$ and
 $ b'' /b =4\m^2 -4\m \d (y)$. Prime is derivative over $y$.

Using (\ref{25}), we obtain the estimate of the components of the tensor 
$\b{R}_\m^\n (\b{g})$  
$$
{R}_\m^\n (\b{g})\sim1/t^2  \sim \m^2\f{\r}{\m}.
$$
The components of the tensor $\k^2 \tau_\m^\n$ are of order $\r$. 
Comparing the expressions for $R_\m^\n$ calculated with the exact and approximate 
metrics,  we find that in the period of late cosmology when $\r/\m< 1$  
and $\m^2 > {R}_\m^\n (\b{g}),\,\,\,
 \m> \k^2 \tau_\m^\n$, 
both expressions for the Ricci tensor coincide up to the terms of order 
$\r/\m$, which is the 
precision of our calculations in the period of late cosmology.

Written with the approximate metric, the equations of motion are 
considerably simpler than with the exact metric. 
The (44) and (4$\m$) and contracted $(\m\n )$ components of equations of motion which follow
from the action (\ref{2f}) are (cf. \cite{vol})
\ba
\l{44}
 &{}& D^\m D^\n h_{\m\n} - D^2 h -\f{3 b'}{2 b}h' =0,\\
\l{m4}
&{}& (D_\m h - D^\n h_{\m\n} )' =0\\
\l{con}
&{}& D^\m D^\n h_{\m\n} -\f{5 b'}{2 b}h'-h'' =0
,\ea
where $h=h_{\m\n}\b{g}^{\m\n}$.
From the Eqs. (\ref{44}) and (\ref{con}) it follows that    
$$
h'' +\f{b'}{b}h' =0, 
$$
or $h'(y,x) =C(x)b^{-1}(y)$, where $C(x)$ is an arbitrary function. Because of the 
reflection symmetry $x\leftrightarrow -x$, the metric components $g_{\m\n}(y,x)$ 
and $h_{\m\n}(y,x)$ are even functions of $y$, and thus $h' (y,x)$ is an odd function of $y$. 
Because $b(y)$ is even in $y$, it follows that $C(x)=0$ and $h' (y,x)=0$. We obtain that 
on the equations 
of motion $h(y,x)$ is independent of $y,\,\,\,h(y,x)=h(x)$.

After imposing the gauge conditions $h_{4\m}=0$ (\ref{gc2}), there  remain residual gauge 
transformations $\xi_\m (y,x) = b (y)C_\m (x)$. Under these transformations the trace $h(x)$ 
transforms as
$\d h(x)=\b{g}^{\m\n} ({D}_\m C_\n (x)+{D}_\n C_\m (x) )$, and by a suitable choice of $C_\m (x)$  can be 
transformed to zero.  

From the Eq. (\ref{m4}) with $h=0$ it follows that $D^\m h'_{\m\n}(y,x) =0$, or $D^\m 
h_{\m\n}(y,x)=k_\n (x)$. Using the remaining residual gauge transformations with the functions $C_\m 
(x)$ such that ${D}^\m C_\m =0$,
we can impose the condition $D^\m h_{\m\n}(y,x) =0$.


\section{Eigenmodes and eigenvalues}


In the background of the approximate metric (\ref{ap}), in the gauge
$D^\m h_{\m\n}(y,x) =0,\,\,\,h_\m^\m (y,x)=0$, 
the ($\m\n$) components of the field 
equations for fluctuations are
\be
\l{e1}
 {h}_{\m\n}''  - 4\m^2 h_{\m\n} + b^{-1}(y) D_\r D^\r h_{\m\n}
+\d (y)4\m h_{\m\n}   =0
.\ee
Expanding the functions $h^\m_\n (x,y)$ 
$$
h^\m_\n (x,y) =\sum_m\,\phi_{(m)\n}^\m (x) h_m (y),
$$
where
$$
D_\r D^\r\, \phi_{(m)\n}^\m = m^2_< \phi_{(m)\n}^\m ,
$$
and the eigenfunctions $h^<_m (y)$ are determined from the equation 
\be
\l{en2}
  {h^<_m}'' (y) - 4\m^2 h^<_m (y)  + e^{2\m |y|} m^2_< h^<_m (y)
+\d (y)4\m h^<_m (y) =0.
\ee
In the region $y>\b{y}$  in (\ref{21a}) the term with the increasing exponent
 becomes dominant.
The approximate metric is
\be
\l{7.5}
ds^2 =dy^2 + e^{2\m y}\left(-dt^2 +\f{a^2 (0,t)}{4}\left(\f{\r}{\m}\right)^2
\eta_{ab}dx^a dx^b \right)
.\ee
The equation for the
eigenmodes is
\be
\l{en15}
{h^>_m}'' (y)-4\m^2 h^>_m (y) + e^{-2\m |y|} m^2_> h^>_m (y) =0
\ee
Eqs. (\ref{en2}) and (\ref{en15}) are solved 
in appendix B. We show that the norm of the function $ h^>_m$ is smaller than that  $h^<_m$.
Effectively, this allows to neglect the contribution from the region $y>\b{y}$. 
Solving the equations we neglect $t$ dependence of $\b{y}(t)$ which in the final expressions can be effectively 
sent to infinity. In the radiation-dominated period of late cosmology  $e^{2\m |\b{y}|}< 1$ and
$(d\b{y}/dt)/\m \b{y} \ll 1$.

We obtain the spectrum
\be
\l{en13}
m_n\simeq \m e^{-\m \b{y}}\left(n\pi +\f{\pi}{2}\right)
\ee
and the  normalized eigenmode $h_m (0)$ 
\be
\l{en14}
h_m (0)\simeq (\m e^{-\m \b{y}})^{1/2}
.\ee
For the following we  need the sum $\sum_n h_{m_n}^2 (0)$, where $m_n$ is
determined by (\ref{en13}). Because of a narrow spacing between the levels, we
change summation to integration and obtain
\be
\l{en21}
\sum_n h_{m_n}^2 (0) \simeq \int \f{dm\, e^{\m\b{y}}}{\m\pi}
\m e^{-\m\b{y}}=\int\f{dm}{\pi}
.\ee
The integral (\ref{en21}) is independent of $\b{y}$. The same measure of
integration  was obtained in \cite{lan1}, where the authors used
the graviton modes of the RS2 model without matter,
 in which case the integration over
$y$ extends to infinity and the spectrum is continuous. Similarity of the results
can be traced to the fact that we performed calculations in the period of late
cosmology neglecting the terms of order $O(\r/\m)$ as compared to unity.
This result can be anticipated also from the equation for the eigenmodes which
does not contain matter terms. Time (temperature) dependence of the Universe
enters only through the  $\b{y}$ , and calculations of appendix B are effectively performed by taking the 
limit $\b{y}\rightarrow \infty$.

\section{Production of Kaluza-Klein gravitons}

Let us calculate the rate of  production
of Kaluza-Klein gravitons in interactions of particles in hot matter on the brane in the 
radiation-dominated period.

Production of Kaluza-Klein gravitons in reactions of particles localized on the brane is calculated 
with the interaction Lagrangian
$$
I =\k\int d^4x\,\sqrt{-\b{g}}\,{h}_{\m\n}(0,x)T^{\m\n}(x) ,
$$
where $T^{\m\n}$ is the energy-momentum of particles on the brane.
The Boltzmann equation which determines evolution of the energy density 
$\h{\r}(t)$ 
of matter on the brane is \cite{kolb,heb,lan2}
\footnote{In this section we use
the unnormalized energy density.}
\be
\l{i1}
\f{d\h{\r}}{dt} +4H\h{\r} =-\sum_n\sum_i  \int \f{d^3 p}{(2\pi )^3}
  \int\f{d^3 k_1}{(2\pi )^3 2E_1}\f{d^3 k_2}{(2\pi )^3 2E_2}
f^{i}_1 (E_1 ) f_2^{i} (E_2 ) |M^{i}_n |^2 (2\pi )^4 \d^4 (k_1 +k_2 -p )
.\ee
Here $f^{i}$ are the Bose/Fermi distributions of colliding particles and
$M^{i}_n$ is the amplitude of reaction $\psi^{i} +\b{\psi}^{i}\rightarrow G$, where $\psi$
and $\b{\psi}$ are standard model particles on the brane (vector, spinor, scalar) 
and $G$ is a state of mass $m_n$ from the graviton Kaluza-Klein tower
\footnote{We have written the rhs of (\ref{i1}) without the factor $1/2$ (cf. \cite{lan2}),
becauce in annihilation reactions  $\psi^{i}$ and $\b{\psi}^{i}$ are different particles
\cite{kolb,rub}.}.
The squared amplitude of annihilation into graviton is \cite{heb}
\be
\l{i2}
 |M^{i}_n |^2 =A_i \f{\k^2}{8}{h}_n^2 (0) s^4 ,
\ee
where $A_i = A_s ,\, A_v ,\, A_f =2/3,\,4,\,1 $ for scalars, vectors and fermions
and $s^2 =(k_1 +k_2 )^2$.
The sum over the graviton states is transformed to the integral following 
(\ref{en21}).
After integration over the angular variables of the momenta of 
interacting particles Boltzmann equation takes the form
\ba
\l{i3}
&{}&\f{d\h{\r}}{dt} +4H\h{\r} 
\\\nonumber
&{}& = -\f{\k^2\sum_i A_i}{8 (2\pi )^4}\int\,
\f{dm e^{\m\b{y}}}{\m}
 {h}_m^2 (0)\, m^4\,\int\, dk_1 dk_2 f(k_1 ) f(k_2 )
(k_1 +k_2 )\theta \left(1-\bigg|1-\f{m^2}{2k_1 k_2}\bigg|\right)
.\ea
Integrating over $m$, we obtain 
\ba
\l{i4}
&{}&\f{d\h{\r}}{dt} +4H\h{\r} = -\f{\k^2 \sum_i A_i}{ 8 (2\pi )^4}
\int\limits_0^{(2k_1 k_2 )^{1/2}}\,dm \,m^4\int\, dk_1 dk_2 f(k_1 ) f(k_2 )
(k_1 +k_2 )
\\\nonumber
&{}&=- \f{\k^2\, A \,315  \zeta(9/2)\zeta(7/2) \,T^8}{\pi^3 2^8},
\ea
where the sum extends over relativistic
degrees of freedom 
\be
\l{i5}
A=\sum_i A_i = \f{2g_s}{3} +g_f (1-2^{-7/2})(1-2^{-9/2})  +4g_v
.\ee
In the high-energy period, when all the standard model degrees of freedom are 
relativistic, $A=166,2$. Eq.(\ref{i4}) has the same form as Eq.(\ref{22a}) and 
the 
expression in the rhs of (\ref{i4}) is the component $\c{T}_{04}$ of the bulk 
energy-momentum tensor \cite{lan2} (see (\ref{i6}) in the next section). The 
physical content of both equations is that the energy loss on the brane 
is due to graviton production.


\section{The energy-momentum tensor in the bulk}

Corrections to the leading-order solution to the Einstein equation are 
expressed through the components of the energy-momentum tensor of the emitted 
gravitons in the bulk. In this section we determine the $y$-dependence of the 
components of the energy-momentum tensor in the bulk in the period of late 
cosmology.

After being produced in reactions on the brane, gravitons move freely 
following the null geodesics in the bulk. Coordinates of particles on the geodesics are 
parametrized by  an affine parameter $\la$. 
In the region $0<y<\b{y}$ we  integrate geodesic equations using
 the approximate metric (\ref{ap})
$$
ds^2 = dy^2 +e^{-2\m y}\b{g}_{\m\n}(x)dx^\m dx^\n
.$$
Geodesic equations are
\ba
\l{g1}
&{}&\f{d^2x^a}{d\la^2}
+2\Gamma^a_{by}\f{dx^b}{d\la}\f{dy}{d\la}+
2\Gamma^a_{b {0}}\f{dx^b}{d\la}\f{dx^0}{d\la}=0\\
\l{g2}
&{}&\f{d^2 x^0}{d\la^2}
+2\Gamma^{0}_{{0} y}\f{dx^0}{d\la}\f{dy}{d\la}+
2\Gamma^{0}_{ab}\f{dx^a}{d\la}\f{d x^b}{d\la}=0\\
\l{g3}
&{}&\f{d^2 y}{d\la^2}
+\Gamma^y_{{0} {0}}\left(\f{dx^0}{d\la}\right)^2+
\Gamma^{y}_{ab}\f{dx^a}{d\la}\f{d x^b}{d\la}=0
,\ea
where connections are
$$
\Gamma^a_{by}=-\d^a_b\m,\quad \Gamma^a_{b
{0}}=\d^a_b\f{\dot{a}}{a},\quad
\Gamma^{0}_{{0} y}=-\m,\quad \Gamma^{0}_{ab}=\eta_{ab}\dot{a}a,\quad
\Gamma^y_{{0} {0}}=-\m e^{-2\m y},\quad \Gamma^{y}_{ab}= a^2\eta_{ab}\m
e^{-2\m y}.
$$
Integrating the geodesic equations, we obtain
\ba
\l{g4}
&{}& \f{dx^a}{d\la}= \f{C^a e^{2\m y}}{a^2(0,t )}\\
\l{g5}
&{}&\f{dt}{d\la}= e^{2\m y}\left({C_0}^2 +\f{C_a C^a}{a^2 (0,t 
)}\right)^{1/2}\\
\l{g6}
&{}&\left(\f{dy}{d\la}\right)^2 -{C_0}^2  e^{2\m y}=C_4
.\ea
Using these relations, it is straightforward to verify that
\be
\l{g7}
\left(\f{dy}{d\la}\right)^2 +e^{-2\m y}\left[-\left(\f{dt}{d\la}\right)^2
+a^2\eta_{ab}\f{dx^a}{d\la}\f{dx^b}{d\la}\right]=const
\ee
is independent of $\la$..

The energy-momentum tensor of the bulk gravitons
is taken in the form of free radiation of massless particles \cite{lan2}
\be
\l{g0}
\c{T}^{ij}(x) =\int\,d^5 p \sqrt{-g}\d (p_i p^i )\t{f}(x,p) p^i p^j ,
\ee
where $\t{f}(x,p)$ is the phase space density of the distribution function
of gravitons.
The function $\t{f}(x,p)$ is determined by identifying the  
 interaction term in the rhs of the Boltzmann equation
(\ref{i4}) and $2\c{T}_{04}$ in the $\{04\}$ component of the Einstein equations 
(\ref{22a}) written in the radiation-dominated period, $p=\r/3$,
in terms of the unnormalized energy density $\h{\r}$ \cite{lan2}
$$
\f{d\h{\r}}{dt} +4H \h{\r}=2 \c{T}_{04}
.$$

The components of momentum $p^i$ of gravitons subject to condition $p_i p^i =0$
are proportional to the tangent vectors to null geodesics
\be
\l{o1}
p^0= N(y,t) e^{2\m y}\left({C_0}^2 +\f{C_a C^a}{a^2
(0,t)}\right)^{1/2},\quad
p^4= N(y,t) {C_0} e^{\m y} ,\quad
p^a (y) = N(y,t) \f{C^a e^{\m y}}{a(0,t)}
,\ee
where $N(y,t)$ is a normalization factor determined below. 
Integrating the geodesic equations (\ref{g4})-(\ref{g6}) with the initial 
conditions
$$
y(\la =0) =0,\qquad t (\la =0)=t_0, \qquad x^a (\la =0)=x^a  
,$$
we obtain $x^i =x^i (\la ,C^i , t_0 , x^a )$. 
Integrating 
(\ref{g6}), we have
$$
\int^y_0 dy' (C_4 +C^2_0 e^{2\m y'} )^{-1/2}=\la
.$$
From this relation is determined $\la =\la (y,C^i )$. Inverting this relation, 
we obtain $y=y(\la ,C^i )$. 
Because of the initial conditions, $y(0,C^i ,x^a )=0$. 

The phase space density of
non-interacting massless particles $\t{f}(x^i , p^i )$ as a function of 
$\la$ is constant along a null geodesic
\be
\l{ge}
\f{d \t{f}(x^i (\la) , p^i (\la ))}{d\la}= p^k
\f{\pa\,\t{f}}{\pa x^k}-\f{\pa\,\t{f}}{\pa p^k}\Gamma^k_{rs}p^r p^s =0
.\ee
Because the  function 
$\t{f}(x^i (\la ,C^i , t_0 , x^a ),\, p^i (\la ,C^i , t_0 ,
x^a ))\equiv \h{f}(\la, C^i, t_0 ,x^a ) $
is constant along geodesics, we have
$$
\t{f}(x^i (\la ,C^i , t_0 , x^a ),\, p^i (\la ,C^i , t_0 ,
x^a ))=
\t{f}(0,t_0 , x^a, p^i (0, C^i ,t_0 ,x^a ))\equiv f(C^i ,t_0 ,x^a ).
$$

Let us consider the $\{ 04\}$ component of the energy-momentum tensor
$$
\c{T}^{04}(y,t,x^a )=\int\,d^5 p \sqrt{-g}\d (p_i p^i )f(x,p) p^0 p^4 
.$$
Changing integration variables $p^i$ to  $C^i$, we obtain
\ba
\l{f2}
\c{T}^{04}(y,t)=\int\,d^5 C^i e^{-4\m y}a^3 (0,t)  \left|\f{\pa (p^0 ,p^4 ,p^a 
)}{\pa (C^0 ,C^4 ,C^a )}\right| 
\d (N^2 C_4 ) f(C^i ,t_0 ,x^a )
\\\nonumber 
N^2 (y,t)e^{2\m y}\left({C_0}^2 +\f{C_a C^a}{a^2 (0,t)}\right)^{1/2}
\left({C_0}^2 e^{2\m y} +C_4 \right)^{1/2}
.\ea
The Jacobian of the transformation is 
\be
\l{f3}
\left(\f{Ne^{2\m y}}{a(0,t)}\right)^3 \f{Ne^{2\m y}C_0}{\left({C_0}^2 +{C_a
C^a}/{a^2
(0,t)}\right)^{1/2}}\f{N}{2\left({C_0}^2 e^{2\m y} +C_4 \right)^{1/2}}
.\ee
We traced the distribution function along the geodesics to $\la =0$, i.e. to the 
hypersurface $y=0$. 
Separating the $y$-dependence, we have
\be
\l{f4}
\c{T}^{04}(y,t)=\c{T}^{04}(0,t )\left(\f{N (y,t)}{N(0,t)}\right)^5 
e^{6\m y}
.\ee
Following \cite{lan2}, we
expand the graviton momentum $p^i$ in the orthonormal basis adapted to 
the foliation of the
space-time
\be
\l{i9}
u^i = (1,0, 0,0,0)e^{\m y},\qquad n^i =(0,0,0,0,1),\qquad e^i_a= (0,1_a,0,0 
,0)\f{e^{\m y}}{a(0,t)}
.\ee
We have  
$$
p^i = E u^i +m n^i + q^a e^i_a
,$$
and $p_i p^i = -E^2 +m^2 +q_a q^a .$

Projections of the components
of the bulk energy-momentum tensor on the plane $y=0$ 
\be
\l{i7}
\c{T}_{\b{0}\b{4}}\equiv \c{T}^{ij}u_i n_j ,\quad
\c{T}_{\b{0}\b{0}}\equiv \c{T}^{ij}u_i u_j,\quad
\c{T}_{\b{4}\b{4}}\equiv \c{T}^{ij}n_i n_j
\ee
have the form \cite{lan2}
\footnote{Cf. (\ref{i4}).}
\ba
\l{i6}
&{}&
\c{T}_{\b{0}\b{4}}(0,t) =-\f{315\, A\zeta (9/2)\zeta (7/2)}{2^{9} 
\pi^3}\k^2 T^8,
\quad \c{T}_{\b{4}\b{4}}(0,t)= \f{3A\zeta(9/2)\zeta(7/2)}{2\pi^4}\k^2 
T^8,\\\nonumber
&{}&\c{T}_{\b{0}\b{0}}(0,t) =\f{21\,A\zeta(9/2)\zeta(7/2)}{8\pi^4}\k^2 T^8
,\ea
where $T$ is the temperature of the Universe at the time $t$.

Separating in $\c{T}_{\b{0}\b{4}}$ the $y$-dependence, 
we have
\be
\l{g12}
\c{T}_{\b{0}\b{4}} (y,t) = \left(\f{N (y,t)}{N(0,t)}\right)^5 e^{5\m y} 
\c{T}_{\b{0}\b{4}} 
(0,t).
\ee
The same relations hold for all components:   
$\c{T}_{\b{i}\b{j}}(y,t)=(N(y,t)e^{\m y}/N(0,t))^5 \c{T}_{\b{i}\b{j}}(0,t) $.

Our next aim is to determine the normalization factor $N(y,t)$.
Lacking an independent definition of momentum of massless particle which could 
enable us
to fix the normalization factor, let us consider the 
energy-momentum tensor conservation equations in the bulk $\nabla_i 
\c{T}^{ij}(y,t)=0$
and try to extract from these equations an information on $y$-dependence of the
energy-momentum tensor.
   
The component $j=0$ of the conservation equations is
\be
\l{g10}
\nabla_i \c{T}^{i0}(y,t)= \pa_4 \c{T}^{40} +(3\G^0_{04}+\G^a_{a4}) \c{T}^{40}+
\pa_0 \c{T}^{00}+ (2\G^0_{00}+\G^a_{a0}) \c{T}^{00}+\G^0_{ab}\c{T}^{ab}=0.
\ee
Substituting the explicit  expressions for the Christoffel symbols, we obtain
\be
\l{g11}
\nabla_i \c{T}^{i0}(y,t)= \pa_4 \c{T}^{40} +\left(3 \f{n'}{n}+3 \f{a'}{a}\right) 
\c{T}^{40}+
\pa_0 \c{T}^{00}+ \left(2\f{\dot{n}}{n}+3\f{\dot{a}}{a}\right) 
\c{T}^{00}+\f{\dot{a}a}{n^2}\eta_{ab}\c{T}^{ab}=0
,\ee
where
\be
\l{g11a}
\c{T}^{04}e^{-\m y} =\c{T}_{\b{0}\b{4}},\qquad 
\c{T}^{00}e^{-2\m y} =\c{T}_{\b{0}\b{0}},\qquad
\c{T}^{44} =\c{T}_{\b{4}\b{4}},\qquad
\c{T}^{ab}e^{-2\m y}a^{-2} (0,t) =\c{T}_{\b{a}\b{b}}.
\ee
In the region $0<y<\b{y}$, using the approximate metric (\ref{24}), we have 
$$
n'/n \simeq -\m, \qquad  a'/a \simeq -\m,\qquad
\dot{a}/a\simeq H(t),\qquad \dot{n}/{n}\sim H(t)\r/\m
.$$ 
Taking 
into account that in the radiation-dominated period $T\sim a(0,t)^{-1}$,  we
obtain that 
$\pa_0 \c{T}_{\b{0}\b{0}}(0,t)\simeq 
-8H(t)\c{T}_{\b{0}\b{0}}(0,t)$.
\footnote{We neglected a weak $T$ dependence of $g_* (T)$.} 
Thus, Eq.(\ref{g11}) can be written as
\be
\l{g12a}
(\pa_y \psi -6\m \psi )\c{T}_{\b{0}\b{4}}(0,t) +He^{\m y}\psi (-5 
\c{T}_{\b{0}\b{0}}(0,t)+\eta_{ab}\c{T}_{\b{a}\b{b}}(0,t))=0
,\ee
where $\psi \equiv e^{\m y}(e^{\m y}N(y,t)/N(0,t) )^5$.  
The trace of the energy-momentum tensor of massless particles is zero,
$\c{T}^i_i =0$.
Using  (\ref{i7}),  we obtain a relation
$$
\c{T}_{\b{a}\b{b}}\eta_{ab}=\c{T}_{\b{0}\b{0}} - 
\c{T}_{\b{4}\b{4}}.
$$ 
Eq.(\ref{g12a}) can be rewritten as 
\be
\l{g13}
\pa_y \psi -6\m \psi  +\psi He^{\m y}B=0
,\ee
where
$
B=(4\c{T}_{\b{0}\b{0}}+\c{T}_{\b{4}\b{4}})/(-\c{T}_{\b{0}\b{4}})\simeq 6.44. 
$
 Solving (\ref{g13}), we obtain
\be
\l{g14}
\left(\f{N (y,t)}{N(0,t)}\right)^5 = \exp\left\{-\f{BH (t)}{\m} (e^{\m y} 
-1)\right\}
.\ee

Next, let us consider the $j=4$ component of the conservation equation  of the 
energy-momentum tensor
\ba
\l{g15}\nonumber
\hspace*{-1cm}\nabla_i \c{T}^{i4}(y,t)= \pa_4 \c{T}^{44 } +(\G^0_{04}+\G^a_{a4}) 
\c{T}^{44}+ \pa_0 
\c{T}^{04 }+(\G^0_{00}+\G^a_{a0}) \c{T}^{04}+ 
\G^4_{00}\c{T}^{00} +\G^4_{ab}\c{T}^{ab}\\ 
\hspace*{-0.5cm}
=\pa_4 \c{T}^{44 } +\left(\f{n'}{n}+3\f{a'}{a}\right)\c{T}^{44} + \pa_0 \c{T}^{04 
}+\left(\f{\dot{n}}{n} +3\f{\dot{a}}{a}\right)\c{T}^{04}+
n'n\c{T}^{00}-a'a \eta_{ab}\c{T}^{ab}=0
\ea 
Repeating the same steps as for the component $j=0$, we obtain
\be
\l{g16}
\pa_y\varphi -5\m\varphi -\varphi He^{\m y}\f{5\c{T}_{\b{0}\b{4}}(0,t)}
{\c{T}_{\b{4}\b{4}}(0,t)}=0
,\ee
where $\varphi \equiv (N (y,t) e^{\m y}/N(0,t) )^5$ and 
$B'= -5\c{T}_{\b{0}\b{4}}/\c{T}_{\b{4}\b{4}}\simeq 6.21$. We have used that at 
$y=0$ there is a relation
$$
n'n\c{T}^{00}-a'a \eta_{ab}\c{T}^{ab}=-\m 
(\c{T}_{\b{0}\b{0}}-\eta_{ab}\c{T}_{\b{a}\b{b}})=-\m \c{T}_{\b{4}\b{4}}
.$$
Solving  (\ref{g15}), we have
\be
\l{g17}
\left(\f{N (y,t)}{N(0,t)}\right)^5 = \exp\left\{-\f{B' H(t)}{\m} (e^{\m y} 
-1)\right\}
\ee
It is remarkable that the constants $B$ and $B'$ which appear in different 
equations are close to each other. The small difference between the constants 
can 
be attributed to approximations to the exact equations. 

\section{ Einstein equations with the bulk energy-momentum tensor}


Having the expressions for the bulk energy-momentum tensor $\c{T}^{ij}$, we are in
position to find the range of $t$ and $y$ within which  solutions of the 
Einstein equations with the bulk energy-momentum tensor included are 
sufficiently close
to the leading-order solutions, i.e. can be considered as perturbations. 
In particular, in Sect.4 we determined the
spectrum of Kaluza-Klein tower of gravitons in the leading approximation which 
is possible, if $\r/\m \ll 1$ and if corrections to the leading approximation 
are small.

Let us consider the $({}^0_0)$ component of the Einstein equations in the bulk 
\be
\l{00}
\f{a'' (y,t) }{a(y,t)} +\f{{a'}^2 (y,t)}{a^2 
(y,t)}-\f{\dot{a}^2 (y,t)}{n^2 (y,t) a^2 (y,t)} 
=2\m^2 +\f{2\k^2}{3} \c{T}^0_0 (y,t)
\ee
The ratio of the bulk term to $2\m^2$ is
\be
\l{f8}
\f{\k^2 \c{T}^0_0 (y,t)}{3\m^2} = -\f{\k^2\c{T}_{\b{0}\b{0}}(y,t)}{3\m^2}= 
\left(\f{N(y,t)}{N(0,t)}\right)^5 (y)e^{5\m 
y}\f{\k^2\c{T}_{\b{0}\b{0}}(0,t)}{3\m^2}
\ee
where
\be
\l{f5}
\f{\k^2\c{T}_{\b{0}\b{0}}(0,t)}{3\m^2}=\f{21 
A\zeta(9/2)\zeta(7/2)}{24\pi^4}\left(\f{180}{g_* 
\pi^2}\right)^2\left(\f{\r}{\m}\right)^2\equiv 
K\left(\f{\r}{\m}\right)^2 .
\ee
Here we have used (\ref{2ca}) to write 
$$
\k^2 T^4 =\f{180 \r}{g_* (T) \pi^2},
$$
where
$g_* = g_s +g_v +7 g_f /8$.
The coefficient $K$ varies in the range from $K\simeq 0.2$ at
characteristic energies of the nucleosynthesis, $T\sim 10^{-(3\div 4)}$ GeV, 
to $K=0.025$  at the energies $\sim 10^{2\div 3} GeV$ at which all the standard 
model degrees of freedom are relativistic.
In this range the ratio $\r/\m$ changes  from $10^{-22\div 29}$ to $10^{(-3\div 
1)}$.
Expression (\ref{f8}) can be written as
\be
\l{f9}
\f{\k^2 \c{T}^0_0 (y,t)}{3\m^2}=K\left(\f{\r}{\m}\right)^2 e^{5\m y}
\exp{\left\{-\f{HB}{\m}(e^{\m y}-1)\right\} }
,\ee
where $ HB/\m\simeq 9.1 (\r/\m)^{1/2}$.
In the region $y<\b{y}$, where $e^{2\m\b{y}}=1+2\m/\r$, for $\m/\r \ll 1$, we 
have
$$
\f{\k^2 \c{T}^0_0 (y,t)}{3\m^2}<C K\left(\f{\r}{\m}\right)^2
,$$
where $C=O(1)$. Fixing $t_l$ so that $\r(t_l)/\m\ll 1$, the ratio (\ref{f8}) can be made 
arbitrary small.

Integrating Eq.(\ref{5}), we obtain
the (04) component of the Einstein equations as
\be
\l{f10}
\f{\dot{a}(y,t)}{n(y,t)}=\dot{a}(0,t)\exp {\left(-\int_0^y \,\f{\k^2
\c{T}_{04}a}{3\dot{a}}dy'\right)}
.\ee
Using the formulas of Sect.7, we have
\ba
\l{f12}
\int_0^y \,\f{\k^2\c{T}_{04}(y' ,t)a}{3\dot{a}}dy'=
\int_0^y \,\f{\k^2\c{T}_{\b{0}\b{4}}(y' ,t)e^{-\m y'}}{3H}dy'
\\\nonumber =
\f{\k^2\c{T}_{\b{0}\b{4}}(0,t)}{3H}\int_0^y \,e^{4\m y'}
\exp\left\{-\f{HB}{\m}(e^{\m y'}-1)\right\}dy'
.\ea
In the period of late cosmology, using the approximate metric 
(\ref{24}), we can set  $\dot{a}(y' ,t)/a(y' ,t)\simeq H(t)$.
The factor at  the integral (\ref{f12}) is
\be
\l{f13a}
\f{\k^2\c{T}_{\b{0}\b{4}}(0,t)}{3H}= \f{315\, A}{3\pi^3 
2^{9}\sqrt{2}}\left(\f{180}{g_*\pi^2}\right)^2\left(\f{\r (t)}{\m}\right)^{3/2} 
\equiv K\left(\f{\r}{\m}\right)^{3/2}
.\ee 
At the temperatures of the Universe of order $10^{-3} GeV$, we have $A=18,\,g_* =10.75$ 
and $K\simeq 0.12$. At the temperatures at which all the Standard model degrees of 
freedom are relativistic, 
$K\simeq 0.011$.  
The integral in (\ref{f12}) is
$$
I=\int_1^{e^{\m y}}dx \,x^3 e^{-b(x-1)}=e^b \f{6}{b^4}(f(b) -f(b e^{\m y} ))
,$$
where $b=HB/\m$ and
$$
f(b) =e^{-b}(1+b +b^2/2! +b^3/3! )
.$$
For $b<1$ approximately $f(b)\simeq 1-b^4 /12$, and for $y<\b{y}$ the integral is 
estimated as 
$$
I\simeq \f{e^b}{2}\left(e^{4\m y} -1\right) 
< 2\f{\r}{\m}\exp\left\{9.1 \left(\f{\r}{\m}\right)^{1/2}\right\}
.$$
For small enough $\r (t_l )/\m$ the expression in the exponent 
in (\ref{f10}) 
\be
\l{f3a}
K\left(\f{\r}{\m}\right)^{5/2}\exp\left\{9.1 \left(\f{\r}{\m}\right)^{1/2}\right\}
\ee
can be made arbitrary small.

To conclude, contribution to the energy-momentum tensor due to graviton 
emission modifies solutions of the Einstein equations, however, 
in the period of late cosmology
 corrections to the leading-order solutions  
 are numerically  small as compared to the leading-order terms.

Let us consider the Friedmann equation (\ref{14a}) with the terms $\c{T}_{ij}$ 
included.
Calculating the first integral in (\ref{14a}), we obtain
\ba
\l{f12a}
&{}&I_{04} =\f{2 \k^2}{3 a^4 (0,t)}\int_{t_l}^t dt'\c{T}_{04}(0,t' )(\m +\r (t' ) ) 
 a^{-4} (0,t')= 
\\\nonumber
&{}&\m\r(t) A_{04}\left(\f{1}{12}\left(\f{1}{\m t_l} -\f{1}{\m t}\right) 
+\f{1}{288}\left(\f{1}{(\m t_l 
)^3}-\f{1}{(\m t)^3}\right)\right)
,\ea
where we have substituted (\ref{i6})
\be
\l{f14a}
\k^2 \c{T}_{04}= A_{04}\r^2 =-\f{315\, A\,\zeta (9/2)\zeta (7/2)}{2^{9} 
\pi^3}\left(\f{180}{g_* \pi^2}\right)^2\r^2 
.\ee
The integral  (\ref{f12a}) has a strong dependence on the value of 
the lower limit.
Therefore, in the integrand
the slowly varying functions can be taken at the times when all the 
Standard
model degrees of freedom are relativistic.
In this period $A=166.2, \quad g_* (T)=106.7$ and $ A_{04}\simeq- 0.126 $.
In the leading approximation $\r (t) \simeq 1/8\m t^2$ and  $1/\m t_l \simeq (8\r 
(t_l 
)/\m)^{1/2}$
 Taking $\r (t_l )/\m \sim 0.1\div 0.001$  and $\m\sim 10^{-12} GeV$, we have
$1/\m t_l\simeq 0.9\div 0.09$ and $T_l \sim (5.1\div 1.6)\cdot 10^2 GeV$. 
For  $\r (t_l )/\m \sim 0.1$ we obtain
\be
\l{f14}
I_{04}\simeq -2\m\r\cdot 0.0048
.\ee
In the same way, calculating $I_{44}$, we obtain
\be
\l{f18}
I_{44}=\f{2 \k^2}{3 a^4 (0,t)}\int_{t_l}^t dt'\c{T}_{44}(t' )(2\m\r(t' ))^{1/2}a^{-4} (0,t') 
= \f{\m\r (t) A_{44}}{48}\left(\f{1}{ (\m t_l )^2}-\f{1}{(\m t)^2}\right),
\ee
where
$$
\k^2 T_{44} =A_{44}\r^2 =\f{3\, A\,\zeta (9/2)\zeta (7/2)}{4
\pi^4}\left(\f{180}{g_* \pi^2}\right)^2\r^2
\simeq 0.10 \r^2
.$$
This yields
\be
\l{f15}
 I_{44}\simeq 2\m\r\cdot 0.00094.
\ee

\section{Graviton emission in the bulk and nucleosynthesis}
\subsection{Graviton emission in the period of late cosmology}

Let us compare abundances of ${}^4 He$ produced
in primordial nucleosynthesis calculated in the models with and without 
account of the graviton emission in the bulk.
We consider the radiation-dominated period of late cosmology, in which 
production of gravitons is sufficiently intensive. 
In the period of late cosmology the leading approximation of the Friedmann equation
coincides with the standard cosmological model.

We estimate the effect of the graviton emission on nucleosynthesis by 
solving  Friedmann equation perturbatively
 keeping in the generalized Friedmann equation (\ref{14a}) the 
term linear in the radiation energy 
density and the terms containing the bulk energy-momentum tensor. The Weyl radiation 
term and the term quadratic in radiation energy density  
 can be treated perturbatively also. We have
\be
\l{k2}
H^2 \simeq 2\m\r  -I_{04}-I_{44}
,\ee
where the integrals $I_{04}$ and $I_{44}$ were defined in the previous section.
We use also the 5D energy conservation equation (\ref{22a}) which is written as
\be
\l{k21}
\dot{\r} +4H\r =-\f{A_{04}\r^2}{3}
.\ee
Let $\b{\r}$ and $\b{H}$  be 
the energy density and the Hubble function calculated in the model without inclusion in the 
Einstein equations the 
terms with the bulk energy-momentum tensor and the Weyl radiation.
Defining
$$
\r =\b{\r}+\r_1 ,\qquad\,\, H=\b{H} +H_1 ,
$$
where $\b{H}^2 =2\m\b{\r}$ and $\r_1$ and $H_1$ are perturbations, and
separating in (\ref{k2}) and (\ref{k21}) the leading-order terms, we obtain
\ba
\l{k3}
&{}&2\b{H} H_1 =2\m\r_1 -\f{2\b{\r}}{3}\int\,dt'\left[A_{04} (\m +\b{\r} 
)+A_{44}\b{H}\right]\b{\r}(t')\\
\l{k4}
&{}&\dot{\r}_1 +4\b{H}\r_1 +4\b{\r}H_1 =\f{A_{04}\b{\r}^2}{3} 
.\ea
Substituting in (\ref{k3}) $\b{\r}(t) \simeq 1/8\m t^2$ and performing integration, we 
have
\be
\l{k6a}
H_1 (t) \simeq \f{\m}{\b{H}}\r_1 -\f{\b{H}A_{04}}{48\m}
\left[ \f{1}{t_l}-\f{1}{t} +\f{1}{24\m^2}\left(\f{1}{t^3_l}-\f{1}{t^3} \right)\right]
-\f{\b{H}A_{44}}{192\m^2}\left(\f{1}{t^2_l}-\f{1}{t^2} \right)
.\ee
Substituting expression (\ref{k6a}) for $H_1$ in (\ref{k4}) and noting that $\b{H}=1/2t$, we 
obtain
$$
\dot{\r}_1 +\f{3}{t}\r_1  = 
\f{A_{04}}{192\m^2 t^3}\left[\f{1}{t_l} +\f{1}{24\m^2}\left(\f{1}{t^3_l}-\f{1}{t^3}  \right)  
\right]
+ \f{A_{44}}{768\m^3 t^3}\left(\f{1}{t^2_l}-\f{1}{t^2}  \right)
.$$
Solving this equation, we find
\be
\l{k5}
\r_1 (t) = \f{C_1}{t^3} +\f{A_{04}}{192\m^2 t^3}\left[\f{t-t_l}{t_l}  +
\f{1}{24\m^2}\left(\f{t-t_l}{t_l^3} -\f{1}{2} \left(\f{1}{t^2_l}-\f{1}{t^2} 
\right)\right)\right]
+\f{A_{44}}{768\m^3 t^3}\left(\f{t}{t^2_l}-\f{2}{t_l}+\f{1}{t} \right)
\ee
The constant $C_1$ should be determined by sewing solutions of Friedmann equation 
in the 
periods of early and late cosmologies. For a moment we set $C_1 =0$, i.e. look for a 
contribution from the period of late cosmology. For $H_1$ we obtain 
\be
\l{k6} 
H_1 (t) = -\f{A_{04}}{1536\m^3 t^2}\left(\f{1}{t^2_l}-\f{1}{t^2}\right)
+
\f{A_{44}}{192\m^2}\left(-\f{1}{t_l t^2}+\f{1}{t^3}\right) \simeq 
-\b{\r}(t)\left(\f{A_{04}}{192\m^2 t_l^2}+\f{A_{44}}{24 \m t_l}\right).
\ee
Expressions (\ref{k5}) and (\ref{k6}) show that corrections to the leading terms are small, i.e. 
the perturbative approach is justified. Note that the leading term proportional to 
$A_{04}/\m t_l$ was canceled in $H_1$.

The mass fraction of ${}^4 He$ produced in primordial nucleosynthesis is
\cite{kolb,rub}
$$
X_4 =\f{2(n/p)}{(n/p) +1}
$$
where the ratio $n/p$ is taken at the end of nucleosunthesis.
Characteristic  temperature of 
the onset of the period of nucleosynthesis (freezing temperature $T_n$ of the 
reaction $n\leftrightarrow p$) is estimated as the temperature at which the 
reaction rate 
$\sim G_F T^5$ is approximately equal to the Hubble parameter \cite{kolb} 
$$
G_F T_n^5 \sim H.
$$
The difference  of the freezing temperatures in the models with and without the account 
of the graviton emission is 
\be
\l{k7}
\f{\d T_n}{T_n}\simeq \f{H_1}{5\b{H}}\simeq 
\f{1}{5}\sqrt{\f{\b{\r} (t_n )}{2\m}}\left(\f{A_{04}}{192\m^2 t_l^2}+\f{A_{44}}{24 \m 
t_l}\right)
,\ee
where
\be
\l{k7a}
\f{\b{\r}(t_n )}{\m}=\f{8\pi ^3 g_* (T_n )}{180}\f{T_n^4}{\m M^3}.
\ee
Substituting $T_n\sim 10^{-3} GeV$ and $\m M^3 \sim (\m M_{pl})^2 \sim 10^{14} GeV^4$, we find 
that the ratio ${\d T_n}/{T_n}$ is very small.
The equilibrium value of the $n-p$ ratio at the freezing temperature 
$$
\left(\f{n}{p}\right)_n =\exp{\left[-\f{(m_n -m_p )}{T_n}\right]}
$$
is very sensitive to the value of $T_n$. Substituting $\d(n/p)_n=(n/p)\ln(n/p)_n \d T_n 
/T_n $, we obtain variation of $X_4$ under
variation of the freezing temperature
\be
\l{k8}
\d X_4 \simeq \f{2}{( n/p +1 )^2}\ln\left(\f{n}{p}\right) \left(\f{n}{p}\right)_n \f{\d 
T_n}{T_n}
\ee
which is also a very small number.

\subsection{Graviton emission in the period of early cosmology}

Next, we make an estimate of the variation of  $X_4$ due to the graviton 
emission in the bulk in 
the period of early cosmology, in which the $\r^2$ term in the Friedmann  
equation is dominant. For the value of $\m\sim 10^{-12} GeV$, the characteristic temperatures
of this period are above $5\cdot 10^2 GeV$.

To make an estimate of the graviton emission,
we assume that the collision integral in the Boltzmann
equation and the expressions for $\c{T}_{ij}$ obtained for the period of late 
cosmology remain qualitatively valid in the early cosmological period. 
The new phenomenon in the early cosmological period is that some of the emitted gravitons can return to the
brane and be again reflected in the bulk with a different momentum. These gravitons do not contribute to
the component $\c{T}_{04}$, because they are not produced, but reflected, but contribute to the component
$\c{T}_{44}$.
The Friedmann equation contains a new term 
$\c{T}_{44}^{(b)}$ 
representing  the energy-momentum tensor of gravitons bouncing back to the 
brane \cite{heb,lan2} 
\ba
\l{k9}
&{}&H^2 (t) = \r^2 (t) +2\m\r (t) +\m \r_w (t)\\\nonumber
&{}&-\f{2\k^2}{3 a^4 (0,t)}\int_{t_c}^t dt'\left[
\c{T}_{04}(t' )(\r (t') +\m)+\c{T}_{44}(t' )H(t' )-\c{T}_{44}^{(b)}(t' )H(t' )\right] 
a^4 (0,t')
.\ea
Here the initial time  $t_c$ is the time of reheating.
Numerical estimates \cite{lan2} and considerations from  the Vaidya model \cite{lan1} 
suggest  at $t\gg t_c$ the dominant contributions  from the integrals of the components  
$\c{T}_{ij}$ and the integral of
 the  term $\c{T}_{44}^{(b)}$ 
are mutually canceled.

Let again $\b{\r}$ and $\b{H}$ be the energy density of matter on the brane and the 
Hubble function calculated in the model without the graviton emission in the period of early 
cosmology. We define
$$
\r =\b{\r} +\r_2 , \qquad\,\,\, H= \b{H}+H_2 ,
$$ 
where in the period of early cosmology $\b{\r} (t)\simeq 1/4t $, and $\b{H}\simeq \b{\r} (t)$ 
(see (\ref{25}).  
It is assumed that $\r_2$ and $H_2$ are perturbations to the leading terms.
From the Friedmann equation we have
\be
\l{k10}
2\b{H}H_2 \simeq 2\b{\r}\r_2  
-\f{2\b{\r}}{3}\int_{t_c}^t dt'\left[A_{04} (\m +\b{\r}
)+A_{44}\b{H} -\k^2 \c{T}_{44}^{(b)}\b{H} \right]\b{\r}(t').
\ee
The conservation law yields
\be
\l{k11}
\dot{\r}_2 +4\b{H}\r_2 +4\b{\r}H_2 =\f{A_{04}\b{\r}^2}{3}
.\ee
In the same way as in the period of late cosmology, we obtain the equation for $\r_2$
\ba
\l{k12}
&{}&\dot{\r}_2 +\f{2}{t}\r_2 -\f{1}{t}\left[\f{A_{04}}{12}\left(\m 
\ln\f{t}{t_c}+\f{1}{4}\left(  
\f{1}{t_c} -\f{1}{t}\right) \right)
+\f{A_{44}}{48}\left(\f{1}{t_c} -\f{1}{t} 
\right)\right]
\\\nonumber
&{}&+\f{1}{12t}\int_{t_c}^t\k^2\c{T}_{44}^{(b)}(t' ) dt' 
=\f{A_{04}}{48t^2},
\ea
where we substituted  $\b{\r} \simeq 1/4t$.
Integrating (\ref{k12}) with the initial condition $\r_2 (t_c ) =0$, we obtain
\ba
\l{k13}
&{}&\r_2 (t) =\f{A_{04}\m}{24}\left(\ln\f{t}{t_c} -\f{1}{2} +\f{t^2_c}{2t^2}\right)
+\f{A_{04} +A_{44}}{96t_c }\left(1-\f{t^2_c}{t^2}\right) 
-\f{A_{44}}{48}\left(\f{1}{t}-\f{t_c}{t^2}\right)
\\\nonumber
&{}&-\f{1}{12t^2}\int^t_{t_c}\,dy y\int_{t_c}^y\,dx \k^2\c{T}_{44}^{(b)}(x) 
\ea
The time $t_c$ of reheating is estimated for the reheating temperature $T_R\sim 5\cdot 10^6
GeV
$\cite{miel}. From the relation $\b{\r} (t_c )/\m\sim 1/4\m t_c$ we have
$$
\f{1}{\m t_c} \simeq\f{16\pi^3 g_* (T_R ) T_R^4}{90 M^3}\sim 4\cdot 10^3,
$$
where we have put $M\simeq (\m M_{pl}^2 )^{1/3}\sim 10^8 GeV$ and $g_* (T_R )\sim 10^2 $.
It follows that $t_l/t_c\sim 1/\m t_c \sim 10^{16}$.

In (\ref{k13}) there is a large term $(A_{04} +A_{44} )/t_c$.
Omitting the small terms, we have
\be
\l{10.14}
\r_2 (t) \simeq\f{A_{04}\m}{24}\left(\ln\f{t}{t_c} -\f{1}{2} \right)
+\f{A_{04} +A_{44}}{96t_c }
-\f{A_{44}}{48}\f{1}{t}
-\f{1}{12t^2}\int^t_{t_c}\,dy y\int_{t_c}^y\,dx \k^2\c{T}_{44}^{(b)}(x)
\ee
On dimensional grounds the term $\k^2\c{T}_{44}^{(b)}(t)$ at small $t$ has the following structure
\be
\l{10.15}
\k^2\c{T}_{44}^{(b)}(t) =\f{b_2}{t^2} + \f{b_1 \m}{t} + b_0 \m^2 +\cdots
\ee
Performing integration of the last term in (\ref{10.14}) and taking  $t\sim t_l$, we obtain
\be
\l{10.5}
\r_2 (t_l) \simeq \f{A_{04}\m}{24}\left(\ln\f{t_l}{t_c} -\f{1}{2}\right)-
\f{A_{44}}{48 t_l}
+\f{A_{04} +A_{44}}{96t_c }
-\f{b_2}{24t_c} + \f{b_2}{12t_l} +\f{b_1\m}{24}\ln\f{t_l}{t_c}
\ee
Next, we equate $\r_2 (t_l)$ and $C_1/t^3_l$ in (\ref{k5}).  At the times $t\sim t_l$,
where $\m t_l \sim 1$, we have
\be
\l{10.16}
C_1 \m^3 \simeq \m\left(\f{A_{04}}{24}\left(\ln\f{1}{\m t_c}
-\f{1}{2}\right)-\f{A_{44}}{48}\right)
+\f{A_{04} +A_{44}}{96t_c }
-\f{b_2}{24t_c} + \f{b_2\m}{12} +\f{b_1\m}{24}\ln\f{1}{\m t_c}.
\ee
The term $C_1 /t^3$ is
\be
\l{10.6}
\f{C_1}{t^3}\simeq\f{1}{\m^2 t^3} \left[\m\left(\f{A_{04}}{24}\left(\ln\f{1}{\m t_c}
-\f{1}{2}\right)-\f{A_{44}}{48}\right)
+\f{A_{04} +A_{44}}{96t_c }
-\f{b_2}{24t_c} + \f{b_2\m}{12} +\f{b_1\m}{24}\ln\f{1}{\m t_c}
\right].
\ee
In the period of late cosmology the term $C_1/t^3$ generates in $H_1$ an additional contribution
$\Delta H_1\simeq \m\Delta \r_1 /\b{H}$
\be
\l{10.7}
\Delta H_1=
16\b{\r} \left[\m\left(\f{A_{04}}{24}\left(\ln\f{1}{\m t_c}
-\f{1}{2}\right)-\f{A_{44}}{48}\right)
+\f{A_{04} +A_{44}}{96t_c }
-\f{b_2}{24t_c} + \f{b_2\m}{12} +\f{b_1\m}{24}\ln\f{1}{\m t_c}
\right],
\ee
where $\b{\r}(t)\simeq 1/8\m t^2$ is the energy density in the period of late cosmology.

If the term  $(A_{04} +A_{44} )/96t_c \sim 3\cdot 10^{-4}/t_c $ was not canceled,
it would produce in $\r_1$ the contribution
$$
\Delta \r_1 =\f{A_{04}+A_{44}}{96t_c}\f{1}{(\m t_c)^3}
\simeq \f{3\cdot 10^{-4}}{(\m t)^3 t_c}
.$$
From (\ref{k6a}) we would have
\be
\l{10.18}
\f{\Delta H_1}{\b{H}}=\f{\m}{\b{H}^2}\Delta \r_1 \simeq \f{1.2\cdot 10^{-3}}{(\m t_c )(\m t)}
.\ee
At time of the nucleosynthesis
$$
\f{1}{8(\m t_n )^2}\simeq \f{4\pi^3 g_* (T_n )}{90}\f{T^4_n}{(\m M_{pl} )^2}.
$$
For $ T_n \sim 10^{-3} GeV$, we have $\m t_n \sim 10^{12}$.
From (\ref{10.18}) we obtain  $\Delta H_1 (t_n )/\b{H} (t_n )\sim 4 $,
which is too
large a value, and would contradict the experimental data.

Assuming that the large terms in (\ref{10.7}) are canceled, we have
\be
\l{10.17}
\Delta H_1 =\f{2\b{\r}(t_n )}{3}\left[ A_{04}\left(\ln\f{1}{\m t_c}
-\f{1}{2}\right) +\f{A_{04}}{2}+b_1 \ln\f{1}{\m t_c}\right].
\ee
 Because at the period of nucleosynthesis $\Delta H_1/\b{H}\sim (\b{\r}(t_n )/\m )^{1/2}$
is a small number,  contribution from the
early cosmology would result in a small variation of $\d X_4 /X_4$.

\section{Conclusions and discussion}

Calculations of this paper where performed under the restriction to
the period of late
cosmology, when $\r (T)/\m<1$, where $\r (T)$ 
is the normalized radiation 
energy density of matter 
on the brane, $\m$ is the scale of the warping factor in the metric. 
For $\m \sim 10^{-12} GeV$,
which we adopted in this paper, the limiting temperatures of the
Universe at
which the approximation of  late cosmology is valid are of order $T_l
\sim 5\cdot 10^2 GeV$.

  In the period of late cosmology it was possible to make a number of
approximations, which enabled us to obtain the analytic expression for the
energy loss from the brane to the bulk due to graviton emission. 
We calculated the spectrum of gravitons
and found the energy-momentum tensor of emitted gravitons 
in the bulk. The Einstein equations were solved perturbatively, 
taking as the leading-order 
solution that without the graviton emission. Using the expressions for the 
energy-momentum tensor of gravitons in the bulk, we estimated corrections to the 
leading-order 
solution to the Einstein equations and showed that the perturbative approach is justified.
We estimated also corrections to the leading-order terms in 
 the generalized Friedmann equation and showed that in
the late cosmological period they are sufficiently small as compared to the
leading-order terms.

Graviton
emission changes the cosmological evolution of matter on the brane and thus production 
of light elements in primordial nucleosynthesis.
Solving the system of the 
generalized Friedmann and the 5D energy conservation equations, which included 
the terms representing the energy flux from the brane to the bulk, we found  
the difference of the mass fractions 
of  ${}^4 He$ produced in primordial nucleosynthesis calculated in the models with and without the 
graviton emission
$$
\f{\d X_4}{X_4}\sim \a \sqrt{\f{\b{\r} (T_n )}{\m}}
.$$  
Here $\b{\r}$ is is the radiation energy density on the brane in the leading 
approximation, without account of graviton emission, $T_n\sim 10^{-3} GeV$ is the 
freezing temperature of   the reaction 
$p\leftrightarrow n$, and $\a\ll 1$ is a small number. 
The ratio $\r (T_n )/\m \sim 10^{-26}$ 
is a very small number, and thus the difference of abundances of  ${}^4 He$ 
calculated in both models is small. 
Crude estimate of the contribution to $\d X_4/X_4$ from the early
cosmological period based on the assumption that in the Friedmann
equation the large contributions from the terms containing the energy-momentum tensor of 
the emitted gravitons cancel due to the bounce of the emitted 
gravitons back to the brane \cite{lan2}
 again shows that it is a small number.

In papers \cite{heb,lan2} as a measure of energy density loss on the brane due to 
graviton production was introduced the integral
$$
\Omega_{lost} =\int_{t_l}^t dt\f{-\Delta \dot{\h{\r}}}{\h{\r}}
,$$
where $-\Delta \dot{\h{\r}}$ is the collision integral in the rhs
of the Boltzmann Eq.(\ref{i4}) which gives the rate of graviton emission.  
Because in \cite{heb,lan2} and in the present paper the collision integrals are 
identical (up to the factor $1/2$),  expressions for $\Omega_{lost}$ up to the factor $1/2$ 
are the same.
Substituting expressions for $\Delta\dot{\h{\r}}$  and $\h{\r}$, we find
$$
\Omega_{lost} =\f{A_{04}}{24\m}\left(\f{1}{t_l}-\f{1}{t}\right),
$$
where $t_l$ is the limiting time at which approximation of the late cosmology is 
valid, $\m t_l \sim1$, and $A_{04}=-0.126$.  
However, calculating  correction $ H_1 (t)$ to the
leading-order Hubble function $\b{H}(t)=1/2t$, we found that in $H_1 (t)$
the terms $H(t) A_{04}/\m t_l$ are canceled. 
The ratio of the would-be correction
 $ H(t) A_{04}/48\m t_l$ to the actual correction $\r (t)(A_{04}/192\m^2 t_l^2 +A_{44}/24\m t_l)$
is of order $(\r/\m )^{1/2}$. As discussed above, at the temperatures of nucleosynthesis this is a small 
number. 

An alternative approach to calculation of emission of the Kaluza-Klein gravitons from the brane to the bulk was
developed in \cite{durrus,rusdur} and refs. therein. In this scenario there are a visible brane moving in the 
AdS bulk and a static brane. 
The motion of the visible brane is 
determined by evolution of matter on the brane. The moving brane acts as a time-dependent boundary to the 5D bulk
leading to production of gravitons via the Casimir effect. Although this picture in appearence is very different 
from that of the present
work, many of the intermediate formulas and final expressions are rather similar. In particular, the energy density 
of the
emitted Kaluza-Klein gravitons on the brane was calculated to be $\r_{KK}\sim 1/a^6$, where $a(t)$ is the scale
factor
in the induced metric on the brane. 
The energy density loss on the brane is $d{\r}_{KK}/dt \sim \dot{a}/a^7$. In the radiation-dominated period of late 
cosmology, using the Friedmann eqation $(\dot{a}/a)^2\simeq 2\m\r$, it is obtained that $T\sim const/a(t)$ 
\cite{kolb,rub} 
and    
$d{\r}_{KK}/dt \sim T^8$ in correspondence with the present work and
\cite{heb,lan1,lan2}.

Graviton emission in the bulk was estimated also in the framework of 
DGP-like models. In the model with the action \cite{DGKN2}
$$
S=M^3\int d^4 x \int_0^R\,dy \sqrt{-g^{(5)}}R^{(5)} +M^3 r_c \int d^4 
x\sqrt{-g^{(4)}}R^{(4)} 
$$
where $M$ is in the $TeV$ range, $M^2_{pl}=M^3 (R+r_c )$,  
 the ratio of the rate of the graviton emission in the 
bulk to the cooling rate of matter
due to cosmological expansion is
$$
\f{d\r}{dt}\bigg|_{em}\bigg/\f{d\r}{dt}\bigg|_{exp}\sim \f{T}{M_{pl}}\a ,
$$
 where $\a \sim 10^{-4}$.

In \cite{DGKN3} was proposed a model with the action
$$
S=M_*^3\int d^4 x dy \sqrt{-g^{(5)}}R^{(5)} +M^2_{pl} \int d^4
x\sqrt{-g^{(4)}}R^{(4)}+\int d^4 x \sqrt{-g^{(4)}}L_{SM}
,$$
  where  $M_*\sim 10^{-12} GeV$.   
The strong gravitational coupling, $1/M_*^2$, presumably, is renormalized by the loops of the standard 
model fields down to $1/M^2_{pl}$. 
The ratio of the rate of the graviton emission to the cooling rate of matter was estimated as
$$
\f{d\r}{dt}\bigg|_{em}\bigg/\f{d\r}{dt}\bigg|_{exp}\sim \f{T^3}{M_{pl}M^2_*}.
$$
This ratio is less than unity for $T< 10^2 GeV$.


 \appendix 
\section{ 3D brane in 5D bulk}

From the system of 5D Einstein equations and junction conditions follows
 the 4D form of the Einstein equations
$$
G^{(4)}_{\m\n}=\k^2_4 T^{(4)}_{\m\n}
,$$
where the Einstein tensor $G^{(4)}_{\m\n}$ is formed with the 4D metric
$g_{\m\n}$, and $\k^2_4 =\k^2\m$.
Explicitly \cite{maeda}
\be
\l{23a}
\k^2_4 T^{(4)}_{\m\n}  =\m\k^2 \tau_{\m\n}+\k^4 \pi_{\m\n} +
\f{2\k^2}{3}\left(\c{T}_{\m\n}+\left(\c{T}_{44}-\f{1}{4}\c{T}^i_i\right)g_{\m\n}\right)
-E_{\m\n}
,\ee
where we have set $\s =\m$.
 Here $\pi_{\m\n}$ is quadratic in $\tau_{\m\n}$, and
$E_{\m\n}$ is the 'electric' part of the 5D Weyl tensor \cite{maeda, maart}.
The component  $T^{(4)}{}_0^0 $  is
\be
\l{24a}
\k^2_4 T^{(4)}{}_0^0 =  -6\m \r -3\r^2 - \m\r_D
,\ee
where
$$
\m\r_D =-\f{2\k^2}{3}\left(\c{T}^0_0
+\c{T}_{44}-\f{\c{T}}{4}\right)-E^0_0 .
$$
Noting that $G^{(4)}{}_0^0 =-3H^2$, we obtain the generalized Friedmann
equation
\be
\l{28a}
H^2 = 2\m \r +\r^2 +\f{\m\r_D}{3}
 .\ee

From the 4D Bianchi identity $D_\m G^{(4)}{}_\n^\m =0$
follows the 4D conservation law $D_\m T^{(4)}{}_\n^\m =0$.  The zero
component of the conservation law, $D_\m T^{(4)}{}_0^\m =0$, is
\be
\l{26a}
\dot{T}^{(4)}{}_0^0 +4H {T}^{(4)}{}_0^0 -H {T}^{(4)}{}_\m^\m =0
.\ee
Substituting in this equation ${T}^{(4)}{}_0^0$ from (\ref{24a}) and expressing $\dot{\r}$ via
(\ref{22a}), we obtain
 (cf. \cite{tan,lan2})
\be
\l{27a}
\dot{\r}_D +4H\r_D +
\f{2\k^2}{\m}\left[\c{T}_{04}\left(\m+\r\right)
+{H}\c{T}_{44}\right]=0
.\ee

\section{Eigenmodes and eigenvalues}

Solution of the equation
\be
\l{e3}
h''_m (y)-4\m^2 h_m (y)  + e^{2\m |y|} m^2 h_m (y)  =0
\ee
is
\be
\l{e4}
h_m (y) =C_1 J_2 (\t{m} e^{\m |y|})+ C_2 N_2 (\t{m} e^{\m |y|})
,\ee
where
$$
\t{m}=\f{m}{\m}.
$$
The term with $\d (y)$ is taken into account by the boundary condition
$$
\left[\f{dh_m (y) }{dy}+\f{2}{\t{m}}  h_m (y)\right]_{y={0_+}}=0
$$
which yields the relation
$$
C_1 J_1 (\t{m}) + C_2 N_1(\t{m}) =0.
$$
Using this relation we obtain the eigenfunctions in the regon $y <\b{y}$ the form
\be
\l{e5}
 h^<_{m_<} (y)=C\left[(N_1(\t{m}) J_2 (\t{m}e^{\m |y|}) -
J_1 (\t{m})N_2 (\t{m}e^{\m |y|})\right].
\ee
\footnote{To simplify the formulas wefrequently omit the sub(superscripts) $>$ and $< $.}
Introducing the functions
$$
f_k (y)=N_1(\t{m})J_k (\t{m}e^{\m |y|}) - J_1 (\t{m})N_k (\t{m}e^{\m |y|})
$$
and using the formula
\be
\l{e6}
\int\,dx x Z_n^2 (ax)=\f{x^2}{2}[Z_n^2 (ax)-Z_{n+1}(ax)Z_{n-1}(ax)]
,\ee
where $Z$ is any linear combination of the Bessel functions, we obtain
\be
\l{e7}
||h^<_m||^2 = 2C^2\int\limits_{0}^{\b{y}} e^{2\m y}f_2^2 (y) dy=
\f{C^2}{\m}\left[e^{2\m \b{y}}\left(f^2_2 ( \b{y})-
f_1 ( \b{y})f_3 ( \b{y})\right)-
\left(f^2_2 (0) -f_1 (0)f_3 (0)\right)\right]
.\ee
Typical masses (energies) of the emitted Kaluza-Klein gravitons
are of order of temperature of the Universe $T$.
In the case of $\m
\sim 10^{-12}$ GeV we have $m/\m\gg 1$. 

For large values of arguments of
the Bessel functions, we obtain
\ba
f_1 (y) \simeq \left(\f{e^{-\m y}}{\pi \t{m}}\right)^{1/2}\left[-N_1 (\t{m})\cos z +J_1
(\t{m})\sin z\right]\\\nonumber
f_2 (y) \simeq \left(\f{e^{-\m y}}{\pi \t{m}}\right)^{1/2}\left[N_1 (\t{m})\sin z +J_1
(\t{m})\cos z\right]\\\nonumber
f_3 (y) \simeq \left(\f{e^{-\m y}}{\pi \t{m}}\right)^{1/2}\left[N_1 (\t{m})\cos z -J_1
(\t{m})\sin z\right],\\\nonumber
\ea
where $z=\t{m}e^{\m y}-\pi/4$.
For large $\t{m}$, substituting asymptotics of the Bessel functions, we
have
$J_1^2 (\t{m})+N_1^2 (\t{m})\simeq 2/\pi \t{m}$.
At the upper limit of integration we obtain
$$
f^2_2 (\b{y}) -f_1 (\b{y}) f_3 (\b{y})\simeq \left(\f{2}{\pi \t{m}}\right)^2 e^{-\m \b{y}}
.$$
At the lower limit we have
$$
f_2^2 (0)=\left(\f{2}{\pi \t{m}}\right)^2,\qquad f_1 (0) f_3 (0) =0
.$$
Combining the above expressions, we have
\be
\l{e10}
||h^<_m||^2 = \f{C^2}{\m}\left(\f{2}{\pi \t{m}}\right)^2 (e^{\m \b{y}} -1)
.\ee

In the region $y>\b{y}$  in (\ref{21a}) the term with the increasing exponent
 becomes dominant.
 We obtain the equations for the
eigenmodes
\be
\l{e15}
{h^>}''_{m_>} (y)-4\m^2 h^<_m (y) + e^{-2\m |y|} m_>^2 h^>_{m_>} (y) =0
\ee
with solutions
\be
\l{e16}
h_{m_>}^>(y) =\t{C}_1 J_2 (\t{m} e^{-\m y})- \t{C}_2 N_2 (\t{m}e^{-\m y}).
\ee
For large $y$, such that  $\t{m} e^{-\m y}\ll 1$, the function
 $N_2 (\t{m}e^{-\m y})\sim (\t{m}e^{-\m y} )^{-2}$
rapidly increases, and to
have normalizable eigenfunctions we  set
$\t{C}_2 =0$.
The norm of the function $h_{m}^>(y)$ is
\be
\l{e17}
||h^{>}_{m_>}||^2 =\t{C}_1^2 \int\limits_{\b{y}}^\infty e^{-2\m {y}}
J_2^2 (\t{m}e^{-\m {y}})dy =\t{C}^2\f{e^{-2\m \b{y}}}{2\m}
[J_2^2 (\t{m}e^{-\m\b{y}}) - J_1 (\t{m}e^{-\m\b{y}})J_3 (\t{m}e^{-\m\b{y}})]
.\ee
At temperatures $T\gg \m$ at which production of gravitons is sufficiently intensive
$\t{m}e^{-\m \b{y}}\sim (T/\m) (\r/\m )^{1/2}\gg 1$, and we can substitute the asymptotics of the
Bessel functions
\footnote{Taking $T\sim 10^2 GeV$, we obtain from (\ref{2ca}) $\r/\m \sim 10^{-4}$ and $T/\m \sim 10^{10}$.}.

Asymptotic of the eigenfunction  (\ref{e5}) is
\be
\l{e51}
h^>_m (y)=\f{2C}{\pi\t{m}}e^{-\m \b{y}/2}\cos(\t{m}(e^{\m\b{y}}-1))
.\ee
Instead of sewing the oscillating functions $h_m (y)$ and $h_{m}^> (y)$, we sew the envelopes
of their asymptotics
\be
\l{e18}
\f{2C}{\pi\t{m}}e^{-\m \b{y}/2} = \t{C}_1 \left(\f{2 e^{\m \b{y}}}{\pi \t{m}}\right)^{1/2}
\ee
giving
$$
\t{C}_1 = C \left(\f{2 }{\pi \t{m}}\right)^{1/2}e^{-\m \b{y}}
$$
Using this relation, we obtain
\be
\l{e20}
||h^{>}_m||^2 \simeq\f{\t{C}_1^2}{\m}e^{-2\m \b{y}}\f{2e^{\m \b{y}}}{\pi\t{m}}=\f{C^2}{2\m}\left(\f{2 }{\pi
\t{m}}\right)^2 e^{-\m \b{y}}
.\ee
Because $e^{\m \b{y}}>1$, the norm (\ref{e20})
 is  smaller than (\ref{e10}).

Effectively, we neglect the contribution from the region $y>\b{y}$ and impose the condition
$h_m (\b{y})=0$.
Substituting $h_m (\b{y})$ from (\ref{e51}), we have
$$
\cos \left(\t{m}(e^{\m \b{y}} -1)\right)=0.
$$
From this equation we obtain the spectrum
\be
\l{e13}
m_n\simeq \m e^{-\m \b{y}}\left(n\pi +\f{\pi}{2}\right).
\ee
Normalization constant $C$ determined from condition $||h_m ||=1$ with the norm (\ref{e10}) is
\be
\l{no}
C\simeq \f{\pi \t{m}}{2}\m^{1/2}e^{-\m \b{y}/2}.
\ee
The normalized eigenmode $h_m (0)$ is
\be
\l{e14}
h_m (0)=\f{2C}{\pi \t{m}}\simeq (\m e^{-\m \b{y}})^{1/2}
.\ee

{\large\bf Acknowledgments}

I would like to thank my colleagues at the Theoretical department of the Skobeltsyn institute for
helpful discussions.


\end{document}